\documentclass[11pt,a4paper]{article}

\def\ra{\rightarrow}
\def\Ra{\Rightarrow}
\def\ra{\rightarrow}

\def\({\left(}
\def\){\right)}
\def\[{\left[}
\def\]{\right]}

\def\a{\alpha}
\def\be{\beta}
\def\g{\gamma}
\def\G{\Gamma}
\def\de{\delta}

\def\la{\lambda}

\def\F{\Phi}
\def\th{\theta}

\def\vf{\varphi}
\def\s{\sigma}
\def\S{\Sigma}

\def\ext{\!\land\!}

\def\Cte{\operatorname{Cte}}

\def\coma{\ ,\quad}

\def\build#1_#2^#3{\mathrel{\mathop{\kern
																			0pt#1}\limits_{#2}^{#3}}}

\def\pe{\!\cdot\!}

\def\dps{\displaystyle}

\def\beq{\begin{equation}}
\def\eeq{\end{equation}}

\def\beqa*{\begin{eqnarray*}}
\def\eeqa*{\end{eqnarray*}}

\def\ba{\hspace{-7mm}\begin{array}{lll}}
\def\ea{\end{array}}

\newcommand\bm[1]{\mbox{\boldmath$#1$}}

\def\ggot{\textfrak{\Large g}}

\def\hgot{\textswab{\large h}}

\usepackage{amsfonts,amsmath,amssymb}
\usepackage{mathrsfs}
\usepackage{amscd}
\usepackage{color}
\usepackage{graphicx}
 
\usepackage{yfonts}

\font\tengot=eufm10  \font\sevengot=eufm7  \font\fivegot=eufm5

\newfam\gotfam
\textfont\gotfam=\tengot
\scriptfont\gotfam=\sevengot
\scriptscriptfont\gotfam=\fivegot

\begin{document}

%\textfrak{\Large This: is: Fraktur ;    g g g ,  h h h , q q q , p p p }

%\textswab{This: is: Schwabacher ;   g g g , h h h , q q q , p p p}

%\textgoth{\Large This: is: Gothic ;   g g g , h h h , q q q , p p p}

\title {SOURCE INTEGRALS OF MULTIPOLE MOMENTS\\ FOR STATIC SPACE--TIMES}

\author{J.L. Hern\'andez--Pastora\thanks{Departamento de Matem\'atica Aplicada. E.T.S. Ingenier\'\i a Industrial de
B\'ejar,  and  IUFFyM. e-mail address: jlhp@usal.es}, J. Mart\' in--Mart\' in\thanks{Instituto Universitario de F\'\i sica Fundamental y Matem\'aticas (IUFFyM)}\, and E. Ruiz\,${}^\dag$ \\
Universidad de Salamanca.  Salamanca,
Spain.  \\}

\date{\today} 

\maketitle

\vspace{-10mm}

\begin{abstract}
The definition of Komar for the mass of a relativistic  source is used as a starting point to  introduce  volume integrals for Relativistic  Multipole Moments (RMM). A certain generalization of the classical Gauss   theorem is used to rewrite  these multipole moments as integrals over a surface at the infinity. Therefore it is shown that the above generalization leads to Asymptotic Relativistic  Multipole Moments (ARMM), recovering the multipoles of Geroch or Thorne,  when  the integrals  are evaluated in asympotically cartesian harmonic coordinates. Relationships regarding the Thorne definition and the classical theory of moments are shown.
\end{abstract}


\large

%%%%%%%%%%%%%%%%%%%%%%%%%%%
%%%%%%%%%%%%%%%%%%%%%%%%%%%%

\section{Introduction}

The early definition of relativistic multipole moments (RMM) for vacuum static asymptotically flat space-time set out in the pioneering work of Geroch \cite{geroch}.  This definition has been extended later by Geroch and Hansen \cite{hansen} to the stationary case. They  have also been defined by Thorne \cite{thorne}, and there are other definitions \cite{beigsimon,quevedo} or approaches \cite{fhp, bakdal} for their calculation. 

\medskip
These multipole moments have been very useful especially in the case of axial symmetry to describe in terms of them some  physical quantities of gravitating sources. They can be used to rewrite observable measurable  physical features of the compact object by means of  test particles orbiting around  that source of the gravi\-tational field in study (precession, geodesic deviation, ISCOs, \dots) \cite{luisgyrv,luisgeodv}. These estimates and calculations have been carried out because vacuum solutions  cons\-tructed in terms of the RMM have been arranged trying to describe space-times  represen\-ting deviations from the spherical symmetry case given by the Schwarzschild  solution  (see for instance \cite{kramer_sol}).

\medskip
In classical gravity multipole moments have a double meaning that allow us to identify these quantities not only with the coefficients of the asymptotic expansion of the gravitational potential but also with the  source of the field through  integral expressions extended over the volume of the gravitating object. There is current interest in resuming work to link the RMM to gravitational sources (in general rela\-tivity), so that we can relate these quantities defined on the outside with physical magnitudes of the stellar object and the source itself. In this sense G\"urlebeck \cite{gurle} gets to express Newtonian multipole moments of the Weyl metrics as a source integral. In the relativistic case we have not even precise definitions that allow us to write RMM as integrals over the  sources, and that is the main objective of this work.

\medskip
The paper is organized as follows. Section 2 is  devoted to show the generalized Gauss theorem (GGT) in classical gravity. Firstly, we remember the  approach of the multipole series of the  gravitational field generated by a compact source, insisting the fact that the asymptotic moments  are also  integrals extended over the volume of the 
stellar object that creates the field. In particular, we restrict ourselves to the case of sources with axial symmetry, which is the usual symmetry in  astronomical objects. In Section 3 the Einstein equations for a static space-time,  domain within  this work shall be restricted, are explained. These equations are the basis of subsequent developments and they are written in two different ways; either by using the
quotient metric related to the time-like Killing congruence or  by using a conformal metric.

\medskip
In Section 4 the definition of Komar  mass \cite{komar} and  its connection 
with Tolman mass \cite{tolman} is addressed. It is also shown that, as is well known,  the result of Komar  for the static case is simply the relativistic version of the Gauss theorem, which serves as the basis for subsequent sections. Section  5 contains the main contribution of this work. By taking as a starting point the  Komar mass,  source integrals of  static multipole moments are proposed. In addition, the structure of  Weyl metrics in harmonic coordinates, previously obtained in \cite{jl_mq,jl_tesis}, can be used to prove that these integrals also provide the   asymptotic Thorne moments in the axially symmetric case. The computation is performed by  using both the quotient metric as well as a conformal metric reaching the conclusion that the use of the last one is more
satisfactory for our purposes. This result should be considered as the relativistic version of the generalized Gauss theorem (RGGT).

\medskip
In Section 6 we revisit the integrals  used by Thorne  to define static multipole moments and show that RGGT can be trivially  applied to obtain  coincidence with the asymptotic moments. Finally, the Appendix is devoted to detail the approximate
expressions obtained in previous works for the   Weyl metric when it is developed in spherical harmonic coordinates,  up to order nine in the inverse of the radial coordinate.

\medskip
Latin indices $i,j,k,\dots$ take values $1,2,3$. Greek ones $\,\a,\be,\la,\mu,\dots$ carry from $0$ to $3$.  Einstein summation rule is used for equal indexes in different positions. The signature of the space-time is $(-,+,+,+)$ and we use relativistic units such that $G=c=1$.

%%%%%%%%%%%%%%%%%%%%%%%%%%%
%%%%%%%%%%%%%%%%%%%%%%%%%%%%

\section{Generalized Gauss theorem in classical gravity}

\subsection{The Poisson equation and the Gauss theorem}

In the newtonian theory the gravitational field  $\vec g$ associated to an isolated compact  stellar object is determined, up to a sign, by the gradient of a potential  $\Phi(\vec x)$ which  satisfies the Poisson equation:
\beq
\Delta\F=4\pi\,\mu(\vec x) \ ,
\eeq
where $\Delta$ represents the ordinary Laplacian operator in three dimensional space and $\mu(\vec x)$ denotes the density of the object, which is a function of the position vector $\vec x$ and it  vanishes outwards from the source.

\medskip
As is well known, an immediate consequence of the equation (1) is the classical Gauss theorem so that,  according to it,  the field flow  throughout the surface boun\-dary $ \partial V $ of any volume $ V $ containing the source is proportional to its mass. Indeed, according to (1) and the definition of the Laplacian operator, the mass $M$  can be written as follows:
\beq
M= \int_V \mu(\vec x)\,d^3\vec x =  \frac1{4\pi}\int_V{\rm div}\,{\rm grad}\F\,d^3\vec x \ ,
\eeq
and taking into account the divergence theorem of Gauss
\beq
\oint_{\partial V} \vec g\pe d\vec \s =- 4\pi M \ ,
\eeq
where $d\vec \s$ is the surface element of the boundary $\partial V$.

\medskip
It should be specified here that  stellar objects  are {\it self-gravitating}, and so  their density not only depends, in general,  on the position  but also on the potential $ \F $ itself. Indeed, let us consider for example a static barotropic perfect fluid object  and consequently with spherical symmetry, that is, 
\beq
\mu=\mu(p) \coma p(r) \coma \mu(r) \coma \Phi(r) \coma r\equiv |\vec x| \ ,
\eeq
$ p $ denoting the pressure.
Now taking into account the Euler  equation
\beq
dp=-\mu(p)\,d\Phi
\eeq
the following conclusion  is obtained by integration:
\beq
p = f(\Phi_\S-\Phi) \quad\Ra\quad \mu = g(\Phi_\S-\Phi)    \ ,
\eeq
i.e., $f$ and $g$ are certain functions of that argument where $\Phi_\S$ is the potential (cons\-tant) on the surface of the star ($p=0$). Therefore we conclude that, in general, the Poisson equation is non-linear \big(see for instance Lane--Emden  equation in  \cite{weinberg}\big). 

%%%%%%%%%%%%%%%%%%%%%%%%%%%%%%%%%%
%%%%%%%%%%%%%%%%%%%%%%%%%%%%%%%%%%

\subsection{ Lagrange--Poisson integral. Multipole series}

The Poisson equation can be reversed thanks to the Green function of the Laplacian operator, allowing us to  rewrite the potential $\F(\vec x)$ by means of the  Lagrange--Poisson integral:
\beq
\Phi(\vec x) = - \int \frac{\mu(\vec y)}{|\vec x - \vec y|}\, \,d^3\vec y  \ ,
\eeq
where the integral extends to all space and having demanded the potential to be null at infinity, so that $\F$  turns out to be at least of class $C^1$.

\medskip
The integral (7) is the basis for the asymptotic multipole development of the potential. Indeed, the Green function supports the following series expansion
\beq
\frac1{|\vec x - \vec y|} = \sum_{n=0}^{\infty}\frac{1}{n!}\left.\frac{\partial \big(1/|\vec x - \vec y\,|\big)}{\partial y^{i_1}\cdots\partial y^{i_n} }\right|_{y=0}\!\!y^{i_1}y^{i_2}\cdots y^{i_n} .
\eeq
Now, it is trivial to see that
\beq
\left.\frac{\partial \big(1/|\vec x - \vec y\,|\big)}{\partial y^{i_1}\cdots\partial y^{i_n} }\right|_{y=0}\!\!\bm{=}\,\,(-1)^n\frac{\partial\, 1/r}{\partial x^{i_1}\cdots\partial x^{i_n} }\,\, \bm{=}\,\, \frac{(2n-1)!!}{r^{n+1}}
n_{i_1i_2\dots i_n} \ ,
\eeq
where
\beq
n_{i_1i_2\dots i_l} \equiv (n_{i_1}n_{i_2}\cdots
n_{i_l})^{\rm TF}
\coma n^i\equiv x^i/r \ .
\eeq
These  indexes are risen and lowered  with the Euclidean metric $\de_{ij}$ and the notation TF (trace free) means {\it traceless part\/} (obviously we are using Cartesian coordinates everywhere). Substituting now (8) and (9) in the integral (7), the following expression  is obtained for the potential:
\beq
\F(\vec x) =-\frac{M}{r} - \sum_{n=1}^\infty \frac{(2n-1)!!}{n!}\frac1{r^{n+1}}
\,M^{i_1\dots i_n}\,n_{i_1\dots i_n} \ ,
\eeq
where
\beq
M^{i_1\dots i_n} \equiv  
\int_V y^{i_1\dots i_n}
\,\mu(\vec y\,)\,d^3\vec y \coma y^{i_1\dots i_n}\equiv \Big(y^{i_1}y^{i_2}\cdots y^{i_n}\Big)^{\rm TF} \ ,
\eeq
$M^{i_1\dots i_n}$ being the  multipole moment of order $n$, which is obviously a completely symmetric object without trace. Consequently, it is clear that it only has $2n + 1$ independent components.
 
\medskip
In view of the above it is important to remember that the multipole moments have a double meaning. On the one hand they are integrals over the source that measure the deviation from the spherical symmetry  and on the other hand they are the coefficients of an asymptotic development in terms of a series expansion.

\medskip
A classic question now arises. Do  multipole moments completely determine the source? The theory of  function moments  ensures that the set of {\it all mathematical moments } of the density $ \mu (\vec x)$ fixes it completely. Nevertheless  the multipole moments $ M ^ {i_1 \dots i_n} $ are completely symmetric  traceless quantities, so they are only a  part of all mathematical  moments. Accordingly, they do not fully determine the source, i.e., many different sources may exist which lead to the same potential (11). It is concluded easily that multipole moments  describe  only the ``skeleton" of the source, i.e., the following distribution density  \cite{tulzy}
\beq
\mu(\vec x) = \sum_{n=0}^\infty\frac{(-1)^n}{n!}
M^{k_1\cdots k_n}\partial_{k_1\cdots k_n}\delta(\vec x) \ ,
\eeq
 $\delta(\vec x)$ being the Dirac delta function and $\partial_k\equiv \partial/\partial x^k$.

%%%%%%%%%%%%%%%%%%%%%%%%%%%%
%%%%%%%%%%%%%%%%%%%%%%%%%%%%

\subsection{Generalized Gauss theorem (GGT)}

The expression (13) shows that the first element of ``skeleton" of the stellar object that creates the gravitational field is simply its mass (monopole moment). On the other hand Gauss theorem states that mass is proportional to the flow of the field (which is, up to a sign, the derivative of the potential) throughout any closed surface containing the star. So naturally the question is whether the remaining elements, i.e., the other multipole moments, can be expressed by means  of flows of some expression  containing the potential and its derivative. The answer lies in what we call {\it generalized Gauss theorem\/}, which is probably well known, but it does not appear in standard texts \footnote {We had knowledge of this theorem thanks to  a preprint of Augusto Espinoza (Universidad Zapoteca). Private communication}. The proof of this theorem is almost as simple as ordinary Gauss theorem proof itself. In fact, it starts from the following identity:
\beq
x^{i_1\dots i_n}\Delta\F = \partial^k\Big[x^{i_1\dots i_n}\,\partial_k\F -
\F\,\partial_k x^{i_1\dots i_n}\Big] + \F \Delta x^{i_1\dots i_n} \ ,
\eeq
and, in addition, it is easy to show the following equation:
\beq
\Delta x^{i_1\dots i_n} = 0 \ ,
\eeq
thus, taking into account the definition of moments (12) as well as the  Poisson equation (1) and the divergence theorem, the desired result is obtained:
\beq
M^{i_1\dots i_n}  =
\frac1{4\pi}\oint\big[x^{i_1\dots i_n}\partial_k\F -
\F\partial_k x^{i_1\dots i_n}\big]d\s^k  \ .
\eeq
That is,  as it happens with the mass, higher-order multipole moments can be expressed as the flow, throughout any surface containing the star, of certain combination of the potential and its gradient.

\medskip
It is important to note that in the above demonstration the Poisson equation  plays an essential role and, in addition,  the definition of moments (12) is a starting point, i.e., they are considered as integrals over the source. It is checked without difficulty that flows (16) coincide with the coefficients (moments) of the  asymptotic multipole expansion (11), whereby the consistency of the result  is ensured.

%%%%%%%%%%%%%%%%%%%%%%
%%%%%%%%%%%%%%%%%%%%%%

\subsection*{$\bullet$  Axial symmetry}

We are going to take interest only in the case where the source of the field possess axial symmetry, which is the usual symmetry of celestial objects. In this case multipole moments have the following structure
\beq
M^{k_1\dots k_n}\equiv M_n\, e^{k_1\dots k_n}\coma e^{k_1\dots k_n}\equiv \big(e^{k_1}e^{k_2}\cdots e^{k_n}\big)^{\rm TF} \ ,
\eeq
where $e^k$ is a unit vector along the positive direction of the  symmetry axis, so that the corresponding moment has a single component $M_n$. Now, considering the formula 
\beq
 e^{k_1\dots k_n}\,n_{k_1\dots k_n} =\frac{n!}{(2n-1)!!} P_n(\cos\th) \ ,
\eeq
where $P_n$ is a Legendre polynomial and $\th$ the polar angle with respect to the symmetry axis, turns out  that the multipole expansion (11) leads to the following expression
\beq
 \F(\vec x) =- \sum_{n=0}^\infty
\frac{M_n}{r^{n+1}}\,P_n(\cos\th) \ .
\eeq
Furthermore, using the formula
\beq
e^{k_1\dots k_n}e_{k_1\dots k_n}= \frac{n!}{(2n-1)!!}  \ ,
\eeq
it is obtained from (17) and (12), the following expression for the moment $M_n$ as a source integral: 
\beq
M_n= \frac{(2n-1)!!}{n!}\, M^{k_1\dots k_n} e_{k_1\dots k_n}=\!\! \int_D\mu(r,\th)\,r^n P_n(\cos\th)\, 
d^3\vec x \coma M_0 \equiv M \ .
\eeq

\medskip
As regards the generalized Gauss theorem (16), it is now written (axial symmetry) as follows
\beq
M_n= \frac1{4\pi}\oint \Big\{r^n P_n(\cos\th)\,\partial_k\Phi - \Phi\,\partial_k\big[r^n P_n(\cos\th)\big]\Big\}d\s^k \ .
\eeq

\noindent
Given the formulas below ($e_{\th}^k $ is the unit vector in the direction of  the meridians):
\beq
\left\{\begin{aligned}
&\partial_k f(r,\th) =\partial_r f \partial_k r + \partial_\th f \partial_k\th = 
\partial_r f \,n_k + \partial_\th f\, \frac1r e_{\th k}
\\[1ex]
&d\s^k = n^k d\s = n^k r^2 \sin\th d\th d\vf
\end{aligned}\right.
\eeq
and using a sphere of radius $ r $ as surface of integration,  the following expression is   obtained:
\beq
M_n = \frac12\, r^{n+1}\!\int_{-1}^{+1} P_n(\cos\th)\,\big(r\partial_r\Phi -n\Phi\big)\,d(\cos\th) \ .
\eeq

\medskip
Verification that (24) reproduces the coefficients of the asymptotic development is trivial. Indeed, taking into account (19) the following is obtained:
\beq
r\partial_r\Phi -n\Phi =  \sum_{n=0}^\infty
\frac{(q+1+n)M_q}{r^{q+1}}\,P_q(\cos\th) \ ,
\eeq
thus, substituting in (24) and recalling the relationship of orthogonality
\beq
\int_{-1}^{+1}P_n(\cos\th) P_q(\cos\th) d(\cos\th) =\frac2{2n+1} \de_{nq} \ ,
\eeq
the desired verification is concluded.

%%%%%%%%%%%%%%%%%%%%%%%%%%%%
%%%%%%%%%%%%%%%%%%%%%%%%%%%%
%%%%%%%%%%%%%%%%%%%%%%%%%%%%

\section{ Static fields. Einstein equations}

As already announced in the Introduction, one of the aims of this work is to look for a possible relativistic extension of the  generalized Gauss theorem in the static case. This means that we will consider a gravitational static compact source, which obviously will generate a static and asymptotically flat  space--time.

\medskip
A space--time of this type is characterized by the existence of an integrable time--like Killing vector field, i.e., it supports orthogonal space--like hypersurfaces (we assume that all of this has a  global character). In an asymptotically cartesian  coordinate system adapted to the  Killing,  the metric is written as follows:
\beq
ds^2 = g_{00}(x^k) dt^2 + g_{ij}(x^k) dx^i dx^j \ ,\quad G=c=1 \ ,
\eeq
with the conditions
\beq
g_{00}\equiv -\xi^2<0\coma \hat g_{ij}(x^k) dx^i dx^j >0 \coma \hat g_{ij}\equiv g_{ij}  \ ,
\eeq
\beq
g_{00} \ra -1+2\frac{m}{r}\coma g_{ij}\ra \de_{ij}+2\frac{m}{r}n_i n_j \ ,
\eeq
where $m$ is the total  mass--energy of the system, $r$ is the associated radial coordinate and $n_i$ is the unit radial vector at infinity.

\medskip
In this case, the Einstein equations for the quotient metric $\hat g_{ij}$ are written as follows \big(see \cite{belgeroch} for instance \big):
\beq
(V_3,\hat g):\ \left\{\begin{aligned}
&\hat\Delta\xi  = 4\pi \xi(-T^0_0 +\hat T)\equiv 4\pi\hat\rho_{\rm to}
\\[1ex]
&\hat R_{ij} - \xi^{-1}\hat\nabla_i\partial_j\xi = 8\pi\left(T_{ij}-\frac12Tg_{ij}\right)
\end{aligned}\right. \ ,
\eeq
where $\hat\nabla_i$ denotes covariant derivative with respect to $\hat g_{ij}$, $\hat\Delta\equiv \hat\nabla_i\hat\nabla^i$ is the Laplacian operator and $\hat R_{ij}$ is the Ricci  tensor. The  sub-index {\it ``to"} by the density $\hat\rho_{\rm to}$ refers to Tolman \cite{tolman}, as we will  justify in the next section. Finally, $T^\a_\be$ is the energy-momentum tensor,  $T\equiv T^\a_\a$  and  $\hat T\equiv T^i_i$ being the traces. In the particular case of a perfect fluid we have that ($u^{\alpha}$ being the quadri-vector of the congruence)
\beq
\left\{\begin{aligned}
T^\a_\be = \rho u^\a u_\be +p\big(\de^\a_\be + u^\a u_\be\big)
\\[1ex]
u^\a = \xi^{-1}\de^\a_{(0)}\coma u_\be = -\xi\,\de_\be^{(0)}
\end{aligned}\right. \ .
\eeq
\beq
\Ra\ \hat\rho_{\rm to} = \xi\Big[+\rho +p(-1+1)+3p\Big] = \xi\big(\rho+3p\big)
\eeq
If one takes into account the Euler equations,  the non--linearity of the Poisson equation for  the self--gravitating case is obviously shown, since the density turns out  to be a non trivial function of the potential $\xi$.

\medskip
Some authors \cite{geroch,thorne,belescard} consider that it is more convenient the use of the conformal metric $ \bar g_{ij} \equiv \xi^2 \hat g_{ij} $ instead of the metric $ \hat g_{ij} $. In the case of the reference \cite{belescard} is argued that $ \bar g_{ij} $ must be regarded as the authentic space metric since the  non-relativistic limit of general relativity should lead not only to the classical  Poisson equation   but also to a flat three-dimensional space.  References \cite {geroch, thorne} are restricted, however, to use $ \bar g_{ij} $ for mathematical convenience. An usual  computation shows that  henceforth the Einstein equations are  written as follows:
\beq
(V_3,\bar g):\ \left\{\begin{aligned}
&\bar\triangle\log\xi  = 
4\pi\xi^{-2} (-T^0_0 +\hat T)\equiv 4\pi \bar\rho_{\rm to}
\\[1ex]
&\bar R_{ij} - 2\partial_i\!\log\xi\,\partial_j\!\log\xi = 8\pi(T_{ij}-\hat Tg_{ij})
\end{aligned}\right. \ .
\eeq
Let us note that  densities $\hat\rho_{\rm to}$ and $\bar\rho_{\rm to}$ are related to each other as follows:
\beq
\bar\rho_{\rm to}\sqrt{\bar g} = \hat\rho_{\rm to}\sqrt{\hat g} \ .
\eeq

We will see later the influence of the use of one or  another metric in what we will call ``relativistic generalized Gauss theorem" (RGGT).

%%%%%%%%%%%%%%%%%%%%%%%%%%%%%%%%%%
%%%%%%%%%%%%%%%%%%%%%%%%%%%%%%%%%%
%%%%%%%%%%%%%%%%%%%%%%%%%%%%%%%%%%

\section{ Mass and flow of Komar}

The quantity  usually called Komar mass is defined in the original article \cite{komar} by means of the following expression:
\beq
M_K = \frac1\chi \int_{\S_3} \!\!\nabla_\la\big(\nabla^\la\xi^\mu - \nabla^\mu\xi^\la\big) d\s_\mu \ ,
\eeq
 $\xi^\la$ being the infinitesimal generator of certain transformation, $\chi =8\pi$ the Einstein constant of gravitation and $d\s_\la$ the normal 1--form to a  three--dimensional surface $\S_3$ of the  space--time considered, i.e,
\beq
d\s_\la = \frac1{3!}\eta_{\la\mu\nu\rho}\,dx^\mu\ext dx^\nu\ext dx^\rho \ ,
\eeq
where $\eta_{\la\mu\nu\rho}$ denotes the element of volume and  $\,\ext\,$ the exterior product. We assume that $ \S_3 $ represents the whole  ordinary three-dimensional space in a certain admissible coordinate system or at least a part of it.

\medskip
Using the covariant Gauss theorem  for the divergence, the  three--dimensional vo\-lume integral (35) can be written as a flow integral throughout the two--dimensional surface $\S_2 = \partial\S_3$. The following result is obtained:
\beq
M_K = -\frac1\chi\oint_{\S_2=\partial\S_3}\!\!\!\! \nabla^\la\xi^\mu\, d\s_{\la\mu} \ ,
\eeq
$d\s_{\la\mu}$ being the 2-form normal to  $\S_2$, i.e.,
\beq
d\s_{\la\mu} = \frac12 \eta_{\la\mu\a\be}\, dx^\a\ext dx^\be \ ,
\eeq
or equivalently, in three-dimensional common language, the normal vector to the surface.

\medskip
We are interested in the case where $ \xi^{\a} $ is a Killing vector field, so we have
\beq
\nabla_\a \xi_\be + \nabla_\be \xi_\a=0\ \Ra\ \nabla_\la\nabla^\la\xi_\mu = -\xi^\rho R_{\rho\mu} \ ,
\eeq
and thus, given Einstein equations, the integral of Komar (35) is written as follows:
\beq
M_K= -2\int_{\S_3}\!\!\left(\xi^\la T_\la^\mu-\frac12 T\xi^\mu\right)d\s_\mu \ .
\eeq
If  $\xi^\a$ is the time-like killing vector of a stationary space--time and we use coordinates adapted to it ($\xi^\a = \de^\a_0$), and the three-surface is  $\S_3: x^0=\Cte$, the following expression is obtained:
\beq
M_K= \int_{\S_3}\!\!\(- T_0^0+ \hat T\)\xi\, d^3\vec x \ ,
\eeq
which coincides with the  Tolman  mass \cite {tolman}, and hence the use of sub--index {\it ``to"} in the density $ \hat \rho_{\rm to} $ (30) is justified. The following notation has been used:
\beq
\hat T \equiv T^i_i\coma \xi \equiv \sqrt{-\xi_\a\xi^\a}= \sqrt{-g_{00}} \coma d^3\vec x\equiv dx^1\ext dx^2\ext dx^3 \ .
\eeq

\medskip
Given the above, when the space-time is generated by a compact object we end up with the following conclusions:

\medskip
{\bf a)} If the  energy--momentum tensor $ T_{\a\be} $ is zero outwards from the stellar object (no electromagnetic field) then the  Komar flow (35) is independent of $ \S_2 $ and in addition it can be expressed as a volume integral there on \big(the  result of  Tolman \cite{tolman}\big)

\medskip
{\bf b)} If $T_{\a\be}$ is {\bf not} zero in the exterior (existence of eletromagnetic field) then the  Komar flow defines the mass by taking the sphere of infinity (assuming that space-time is asymptotically flat and $T_{\a \be} $ goes to zero quickly enough).

\medskip
In order to illustrate these ideas we present below the expression  of Komar mass for the Kerr--Newman metric \cite{kerr_newman} written in  Boyer--Lindquist  coordinates (the computation is annoying, but is a straightforward calculation by  using a computer)
\beq
M_K = m-\frac{e^2}{2r} -\frac{e^2}{2a}\(1+\frac{a^2}{r^2}\)\mbox{arctg}\frac{a}{r} \ ,
\eeq
where $\S_3$ has been considerd to be a regular sphere of radio $r$ (radial coordinate) and the parameters $\{m,a,e\}$ denote, as usual,  mass, angular momentum per unit of mass, and electric charge respectively. Meanwhile $ r $ remains finite both the charge and angular momentum contribute to the Komar  mass. Nevertheless, as we go to infinity ($r\to\infty$)  the following result is obtained:
\beq
M_K = m - \frac{e^2}{r} - \frac{e^2 a^2}{3r^3}\left[1+O\!\(\frac{a}{r}\)\right]\, \ra\, m \ ,
\eeq
i.e. Komar mass is reduced to the parameter $m$.

\medskip
It is worthwhile noticing that, at least in the case of a static space-time
 ($g_{0j}=0$ in coordinates adapted to the time-like Killing vector), the result of Komar is simply the gravitational Gauss theorem in its relativistic version. Indeed, let us remind the first Einstein equation (30):
\beq
\hat\Delta \xi = 4\pi \hat \rho_{\rm to}\coma \hat\rho_{\rm to}\equiv \xi \big(T^0_0-\hat T\big) \ ,
\eeq
which is a Poisson equation with the indicated density. Then it  seems to be natural the use of this  density as bulk mass density, and henceforth the total mass will be written as
\beq
M = \int_V \hat \rho_{\rm to}\, \hat\eta \ ,
\eeq
 $V$ being any compact volume containing the source, and  $ \hat\eta$ being  the three-dimensional volume element, i.e.,
\beq
 \hat\eta = \sqrt{\hat g}\, dx^1\!\ext dx^2\!\ext dx^3  \ .
\eeq
As a result, by applying the divergence theorem in its covariant form, we have that 
\beq
M = \frac1{4\pi}\int_V \hat\Delta\xi\,\hat\eta =  \frac1{4\pi}\int_V \hat\nabla_k\hat\nabla^k\xi\,\hat\eta =  \frac1{4\pi}\oint_{\partial V}\hat\nabla^k\xi\,d\s_k \ ,
\eeq
and it is trivial to see that (48) is precisely the result of Komar (37) for the static case,  $\partial V$ being the boundary and  $d\s_k$ its corresponding surface element:
\beq
d\s_k = \frac12 \hat\eta_{kij} \, dx^i\!\ext dx^j = \sqrt{\hat g}\,\underbrace{\frac12 \epsilon_{kij} \, dx^i\!\ext dx^j}_{\dps \equiv d\tilde\s_k} \ .
\eeq

\noindent We can also apply the divergence theorem in its non-covariant form as follows:
\beq
M = \frac1{4\pi}\int_V \hat\Delta\xi\,\hat\eta =  \frac1{4\pi}\int_V\frac1{\sqrt{\hat g}}\partial_k\big(\sqrt{\hat g}\,\hat g^{kj}\partial_j\xi\big) \sqrt{\hat g}\, d^3\vec{x}
\eeq

\medskip
\beq
\Ra\ M= \frac1{4\pi}\oint_{\partial V}\hat g^{kj}\partial_j\xi\,\sqrt{\hat g}\, n_k\, d\tilde\s \ .
\eeq
%%

%%%%%%%%%%%%%%%%%%%%%%%%%%%%%%%%%%
%%%%%%%%%%%%%%%%%%%%%%%%%%%%%%%%%%
%%%%%%%%%%%%%%%%%%%%%%%%%%%%%%%%%%

\section{Static multipoles  {\it ``\,\`a la Komar\,"}.\\ Relativistic generalized Gauss theorem (RGGT)}

As we  restrict  ourselves to the static case, it is clear from (34), (46) and (48) that Komar mass can be treated  in two different ways: either by using the quotient metric $\hat g_{ij}$ or by using the conformal quotient metric $\bar g_{ij} = \xi^2 \hat g_{ij}$. This is what we will do in this section.

%%%%%%%%%%%%%%%%%%%%%%%%%%%%%%%%%
%%%%%%%%%%%%%%%%%%%%%%%%%%%%%%%%%

\subsection{Quotient metric}

The expression (46) provides the Komar mass of the system, which obviously coincides with the  first  coefficient (monopole) of  the asymptotic development of the metric. In addition, taking into account (48) it is clear that it can be written as an integral over the whole space of a function obtained from the gravitational field. In this way we will generalize (46) and {\it define } the other multipole moments as follows:
\beq
\begin{aligned}
M_{K}^{i_1\dots i_n} &= \int \! x^{i_1\dots i_n}\,\hat\rho_{\rm to}\,\hat\eta -\frac1{4\pi} \int \xi\, \hat\Delta x^{i_1\dots i_n}\,\hat\eta
\\[1ex]
&= \frac1{4\pi}\int \Big[ x^{i_1\dots i_n}\,\hat\Delta\xi -  \xi\, \hat\Delta x^{i_1\dots i_n}\Big]\hat\eta \ ,
\end{aligned}
\eeq
where  $\, x^{i_1\dots i_n}\,$ still being the traceless part of the product  $\,x^{i_1}\!\cdots x^{i_n}\,$ with respect to the euclidean metric $\de_{ij}$, i.e., as an example
\beq
 x^{ij}= x^i x^j -\frac13 \de^{ij} \de_{kl} x^k x^l \ ,
\eeq
in such a way that  $\,M_{K}^{i_1\dots i_n}\,$  turns out to be a completely symmetric object without trace with respect to $\de_{ij}$.  We also require the coordinates used  to be  asymptotically cartesian harmonic coordinates. The ultimate justification of this definition is that, as we will see, these moments coincide, at least in the case of axial symmetry, with asymptotic standard multipole moments of Thorne \cite {thorne} or Geroch \cite {geroch}.

\medskip
The same  process used to prove the  generalized Gauss theorem in classical gravi\-tation  can be used now to express the above definition
as a flow integral. Indeed, given that
\beq
\big[x^{I_{(n)}} \hat\Delta\xi - \xi\hat\Delta x^{I_{(n)}}\big]\sqrt{\hat g} \equiv x^{I_{(n)}}\partial_k\!\left[\!\sqrt{\hat g}\, \hat{g}^{kj} \partial_j \xi\right] -\xi\partial_k\!\left[\!\sqrt{\hat g}\, \hat{g}^{kj} \partial_j x^{I_{(n)}}\!\right] \ ,
\eeq
\big(${I_{(n)}}\equiv i_1i_2 \cdots  i_n$\big) as well as,
\beq
\begin{aligned}
x^{I_{(n)}}\partial_k\!\left[\!\sqrt{\hat g}\, \hat{g}^{kj} \partial_j \xi\right]&=\partial_k\left[x^{I_{(n)}}\sqrt{\hat g}\, \hat{g}^{kj} \partial_j \xi\right] - \sqrt{\hat g}\, \hat{g}^{kj} \partial_j \xi\, \partial_k x^{I_{(n)}}
\\[1ex]
\xi\,\partial_k\!\left[\sqrt{\hat g}\, \hat{g}^{kj}\partial_jx^{I_{(n)}}\right]&=\partial_k\!\left[\xi\sqrt{\hat g}\, \hat{g}^{kj}\partial_jx^{I_{(n)}}\right] - \sqrt{\hat g}\, \hat{g}^{kj} \partial_j \xi\,\partial_kx^{I_{(n)}} \ ,
\end{aligned}
\eeq
it turns out that we can rewrite (52) as follows:
\beq
4\pi M^{I_{(n)}}_K=\int \!\partial_k\!\left[x^{I_{(n)}}\sqrt{\hat g}\, \hat{g}^{kj} \partial_j \xi -\xi\sqrt{\hat g}\, \hat{g}^{kj}\partial_jx^{I_{(n)}}\right]\! d^3 \vec{x} \ ,
\eeq
and once again making use of the divergence theorem, it is finally obtained
\beq
M^{I_{(n)}}_K=\frac1{4\pi}\!\oint_\infty \Big[x^{I_{(n)}}\, \hat{g}^{kj} \partial_j \xi -\xi\, \hat{g}^{kj}\partial_j x^{I_{(n)}}\Big]\sqrt{\hat g}\, d\tilde\sigma_k  \ ,
\eeq
result that can be considered as the {\it relativistic generalized Gauss theorem} (RGGT) for the static case. Let us note that the application of the divergence theorem requires gravitation potentials to be of  class $ C^1 $, which in principle can be ensured by the use of certain harmonic coordinates \cite{mmr}.

\medskip
We shall restrict ourselves again to the  axial--symmetry case. As usual  we  will use a coordinate system adapted to both two  Killing vector fields and we assume that the metric has Papapetrou structure in associated spherical coordinates \cite {mmr}, i.e.,
\beq
ds^2 = g_{tt} dt^2 + g_{rr} dr^2 +2g_{r\th} dr d\th + g_{\th\th} d\th^2 + g_{\vf\vf} d\vf^2 \ ,
\eeq
where the azimuthal coordinate $\vf$ defines the axial symmetry and  hence all metric components are only functions of the  radial coordinate  $r$ and the polar coordinate $\th$. In particular, we  write the metric in the following way:
\beq
ds^2 = \g_{tt}\, dt^2 + \g_{rr}\, dr^2 +2\g_{r\th}\, dr (r d\th) + \g_{\th\th}\, (r d\th)^2 + \g_{\vf\vf}\, (r \sin\th d\vf)^2 \ ,
\eeq
that is, we are using the ``orthonormal euclidean cobasis"
\beq
\{dr, r d\th, r \sin\th\, d\vf\} \ ,
\eeq
so that,
\beq
g_{tt} =\g_{tt}\coma g_{rr}= \g_{rr}\coma g_{r\th}= r\, \g_{r\th}\coma g_{\th\th}=r^2 \g_{\th\th} \coma g_{\vf\vf} =r^2\sin^2\!\th\,\g_{\vf\vf} \ .
\eeq

\medskip
Now considering the above and the results of Section 2 concerning the axial symmetry, a small calculation shows that the flow  integral (57) for axisymmetric multipole moments  {\it ``\,\`a la Komar\,"}  is written as follows:
\beq
\begin{aligned}
M^K_n &= \frac14\oint_{-1}^{+1}\!\!r^{n+1}P_n(w)\,(- g)^{-1/2}\,\g_{\vf\vf}\g_{\th\th}\Big[2n\, \g_{tt} - r\, \partial_r\g_{tt}\Big]dw
\\[1.5ex]
&\quad +\frac14\oint_{-1}^{+1}\!\!r^{n+1}P_n(w)\,(- g)^{-1/2}\,\g_{\vf\vf}\, \g_{r\th}\,\partial_\th\g_{tt}\,dw
 \\[1.5ex]
&\quad -\frac12\oint_{-1}^{+1}\!\!r^{n+1}P_n^1(w)\,(- g)^{-1/2}\,\g_{tt}\, \g_{\vf\vf}\,\g_{r\th}\,dw \ ,
\end{aligned}
\eeq
where $g$ is the determinant of the metric of the space-time in the euclidean basis
\beq
g= \g_{tt}(\g_{rr} \g_{\th\th} - \g_{r\th}^2)\g_{\vf\vf} \ ,
\eeq
 $w\!\equiv \!\cos\th$ being the notation used for the variable of integration. The circle used in the sign of integration means that the sphere of infinity must be considered, i.e., we needs to take  $r\to\infty$. Therefore all terms arising like  $1/r^k$ with $k\ge1$ should be excluded.

\medskip
The calculation of the flow integrals (62) requires the use of the structure (104) of the metric in harmonic coordinates set out in the Appendix. From that, it is easy to obtain the following structures for different terms of (62):
\begin{align}
&\hspace{-.2cm}(-g)^{-1/2}\g_{\vf\vf} \g_{\th\th} \g_{tt}=  -1 + \sum_{q=0}^\infty \frac1{r^{q+2}}X_1^{(q)}(w)
\\[1ex]
&\hspace{-.2cm}(-g)^{-1/2}\g_{\vf\vf} \g_{\th\th}\, r\, \partial_r\g_{tt} =\! -2\sum_{q=0}^\infty \frac{q+1}{r^{q+1}}M_q P_q(w) +\! \sum_{q=2}^\infty \frac1{r^{q+3}}X_2^{(q)}(w)
\end{align}
\begin{align}
&(-g)^{-1/2}\g_{\vf\vf} \g_{r\th}\,\partial_\th\g_{tt} = \sum_{q=0}^\infty \frac1{r^{q+5}}X_3^{(q+2)}(w)
\\[1ex]
&(-g)^{-1/2} \g_{tt} \g_{\vf\vf} \g_{r\th} = \sqrt{1-w^2}\, \sum_{q=1}^\infty \frac1{r^{q+3}}X_4^{(q)}(w) \ ,
\end{align}
$X^{(q)}_a(w)\ (a=1,\dots,4)$ being polynomials of degree  $q$ in the variable $w$.

\medskip
As an example to fix ideas on how to do the  flow integrals we choose  the terms (65) and (67). As regards (65)  the following integrals are evaluated:
\begin{align}
&+\frac12\oint_{-1}^{+1} \!\!r^{n+1} P_n(w)\sum_{q=0}^\infty \frac{q+1}{r^{q+1}}M_q P_q(w)\, dw = \frac{n+1}{2n+1}
\\[1ex]
&-\frac14 \oint_{-1}^{+1} \!\!r^{n+1} P_n(w)  \sum_{q=2}^\infty \frac1{r^{q+3}}X_2^{(q)}(w)\, dw=0 \ .
\end{align}
The result of the first integral above (68) is a consequence of  the orthogonality relation (26) whereas for the integral (69) must be taken into account the inequality $ q + 3 \le n + 1 $, in order to avoid terms like $ 1 / r $ or higher, which vanish at infinity. But this implies that $ q \le n-2 $, which means that the maximum degree of a Legendre polynomial from $ X_2 ^ {(q)} (w) $ is $ n-2 $. Applying again the relation (26) it is obtained that this integral is zero.

\medskip
Let us now analyse the term from (67), which leads to the following  integral:
\beq
 -\frac12\oint_{-1}^{+1}\!\!r^{n+1}P_n^1(w) \sum_{q=1}^\infty \frac1{r^{q+3}}\sqrt{1-w^2}\,X_4^{(q)}(w)\,dw \ .
 \eeq
Inequality $q+3\le n+1$ must  be considered again, which implies that $q+1\le n-1$. However, the product of the factor $ \sqrt {1-w^ 2} $ by the polynomial $ X^{(q)}_4 $ generates associated Legendre functions $ P^1_m $ whose maximum degree is $ m = q + 1 $. Consequently, the integral (70) is zero under the  orthogonality relation
\beq
\int_{-1}^{+1}P^1_n(w) P^1_q(w) dw =\frac{2n(n+1)}{2n+1} \de_{nq} \ .
\eeq

\medskip\noindent
Arguing similarly with the other terms (64) and (66), we can conclude that both of them  lead to null integrals. Consequently the unique  non-vanishing integral is (68), whereby the expression (62) is finally reduced to the following:
\beq
M^K_n = \frac{n+1}{2n+1} M_n \ .
\eeq

It is concluded that the axisymmetric moments {\it ``\,\`a la Komar\,"} are proportional to the known moments of Thorne \cite {thorne} (the proportionality constant depends on the multipole order).

%%%%%%%%%%%%%%%%%%%%%%%%%%%%%%%%%%
%%%%%%%%%%%%%%%%%%%%%%%%%%%%%%%%%%

\subsection{Conformal quotient metric}

The result (72) would be more satisfactory if Komar moments exactly coincide  with  Thorne moments. In this Subsection we will modify the definition (52) to get that matching. To do this we assume that  Komar mass  can also be written respect to  the conformal metric, accordingly with (34), as follows
\beq
M_K = \int \bar \rho_{\rm to}\, \bar \eta
\eeq
Hence, given Einstein equations (33), this equality suggests replacing the definition (52) of the multipole moments for the following one:
\beq
\bar M^{i_1i_2\dots i_n}_K  = \frac1{4\pi}\int \big[x^{i_1i_2\dots i_n} \bar\Delta\log\xi - \log\xi\,\bar\Delta x^{i_1i_2\dots i_n}\big]\bar\eta \ ,
\eeq

\noindent
whereby it follows that, similar to (57),
\beq
\bar M^{I_{(n)}}_K = \frac1{4\pi}\!\oint_\infty\!\Big[x^{I_{(n)}}\,\bar \nabla^l\!\log\xi - \log\xi\, \bar\nabla^l \!x^{I_{(n)}}\Big]\sqrt{\bar g}\,d\tilde\s_l \ ,
\eeq
and restricting ourselves to the axial symmetry case, a standard calculation leads to the following:
\begin{align}
\bar M^K_n &= \frac14\oint_{-1}^{+1}\!\! r^{n+1} P_n(w) (-g)^{-1/2} \g_{\vf\vf} \g_{\th\th}\Big[n \,\g_{tt}\log(-\g_{tt})  -  r\partial_r\g_{tt}\Big]dw \notag
\\[1ex]
&\quad +\frac14 \oint_{-1}^{+1}\!\! r^{n+1} P_n(w) (-g)^{-1/2} \g_{\vf\vf} \g_{r\th}\,\partial_\th\g_{tt}\, dw \notag
\\[1ex]
&\quad-\frac14\oint_{-1}^{+1}\!\! r^{n+1} P^1_n(w) (-g)^{-1/2} \,\g_{\vf\vf}\g_{r\th}\, \g_{tt}\log(-\g_{tt})\, dw \ .
\end{align}

To calculate these integrals we proceed in a  similarly way than  the previous case where we used the metric $ \hat g_ {ij} $. We need structures (65), (66) and the following expressions:
\begin{align}
&\hspace{-.35cm}(-g)^{-1/2}\g_{\vf\vf} \g_{\th\th}\,\g_{tt}\log(-\g_{tt})  =  2\! \sum_{q=0}^\infty \frac{M_q}{r^{q+1}} P_q(w) +\! \sum_{q=0}^\infty \frac1{r^{q+3}}X_5^{(q)}(w)
\\[1ex]
&\hspace{-.35cm}(-g)^{-1/2} \g_{\vf\vf} \g_{r\th}\, \g_{tt}\log(-\g_{tt})= \sqrt{1-w^2}\, \sum_{q=1}^\infty \frac1{r^{q+4}}X_6^{(q)}(w) \ ,
\end{align}
where $X^{(q)}_a(w)\ (a=5,6)$ are again polynomials  of degree $q$ in the variable $w$. This time the unique non-zero integrals  are the following ones:
\beq
\frac12\oint_{-1}^{+1} \!\!r^{n+1} P_n(w)\sum_{q=0}^\infty \frac{q+1}{r^{q+1}}M_q P_q(w)\, dw = \frac{n+1}{2n+1} \ ,
\eeq
\beq
\frac12\oint_{-1}^{+1} \!\! n\, r^{n+1} P_n(w)\sum_{q=0}^\infty \frac1{r^{q+1}}M_q P_q(w)\, dw = \frac{n}{2n+1} \ ,
\eeq
whereby the following result is now obtained:
\beq
\bar M^K_n  = \frac{n}{2n+1}M_n + \frac{n+1}{2n+1} M_n= M_n \ .
\eeq

As we see, the use of the conformal metric  $ \bar g_{ij} $ leads to a more satisfactory result  than the one obtained with the quotient metric. This fact represents an additional argument for the use of the conformal metric  against the quotient metric as a metric of space.

%%%%%%%%%%%%%%%%%%%%%%%%%%%%%%%%
%%%%%%%%%%%%%%%%%%%%%%%%%%%%%%%%

\subsection{Comments}

\hspace{4mm} {\bf A)} Firstly, it should be noted here  that although it may seem somewhat artificial the use of the ``euclidean" factor $x^{i_1i_2\dots i_n}$ in (52) and (74), however we should consider some important items: {\bf 1)} The moments are required to be {\it numerical \ } objects  completely symmetric and {\it without  trace}, and hence this condition should  be verified respect to  a metric independently of any coordinate system. {\bf 2)} The definitions (52) and (74) lead to flow integrals throughout the surface of the {\it infinity\/}, where we assumed that the metric is {\it flat}. {\bf 3)} Henceforth the moments $ \bar M_{_K}^{i_1 \dots i_n} $ \big (as well as $ M_{_K} ^ {i_1 \dots i_n} $ \big) should be considered as tensors in a neighbourhood  of infinity so that everything were consistent.

\bigskip
{\bf B)} Secondly, another relevant feature of the  definitions  (52) and (74) has to do with the requirement imposed on  the calculation  of the integrals in asymptotically cartesian harmonic coordinates. This fact means that the above definitions are not covariant and so they depend on the system of coordinates. Let us see what happens, for example, if we develop the definition (52) in canonical spherical coordinates of Weyl for the static--axisymmetric case [metric (100) at the Appendix]. 

\medskip
In this case the integral (52) is reduced to the following:
\beq
\Big[r^n P_n\,\hat\Delta e^\Psi -e^\Psi \hat\Delta\big(r^n P_n\big)\Big]\sqrt{\hat g} = \partial_k\Big\{\be\big[r^n P_n \partial^k \Psi-\partial^k(r^n P_n)\big]\Big\} \ ,
\eeq
and then the following result is easily obtained:
\beq
M_n^K = \frac1{4\pi}\!\!\oint_{\rm ext}\!\underbrace{\big[r^n P_n\partial^k\Psi -\partial^k(r^n P_n)\big]n_k}_{\dps r^{n-1}P_n\big[r\partial_r\Psi-n\big]}d\tilde \s =-\frac{n+1}{2n+1}\, a_n \ ,
\eeq
which compared to (72) highlights the importance of using harmonic coordinates. What happens if we do this calculation with the conformal quotient metric? A similar calculation leads to the following:
\beq
\bar M_n^K  = -\,a_n \ ,
\eeq
as expected if we remember (72) and (81).

\medskip
{\bf C)} Finally, we must underline  recent works of G\"urlebeck \cite{gurle} who obtain an integral expression,  restricted to the interior of the source, for the Weyl coefficients  $a_n$ (newtonian multipole moments) which apparently is independent of the  coordinate system. He states that this result is equivalent to get $ M_n $ multipole moments, as their relationship with $ a_n $ is known \big (see for example \cite {jl_mq, joseluis} \big). In a next work we will establish a comparison between the result of G\"urlebeck with our one by explicitly  taking  an {\it inner } solution which describes an anisotropic fluid.

%%%%%%%%%%%%%%%%%%%%%%%%%%%%%%%%%%
%%%%%%%%%%%%%%%%%%%%%%%%%%%%%%%%%%
%%%%%%%%%%%%%%%%%%%%%%%%%%%%%%%%%%

\section{Relativistic generalized Gauss theorem {\it ``\,\`a la Thorne\,"}}

In this section  we first recall Thorne approach    \cite {thorne} to define the multipole moments using {\it volume integrals } over all space. From there we show that  the generalized Gauss theorem  can be applied almost identically to that of classical gravitation. It may simply be concluded  that such defined moments coincide with the multipole moments of the asymptotic development used by Thorne itself.

\medskip
As is well known  Einstein equations can be written in terms of the metric density 
$\,\ggot^{\a\be} \equiv \sqrt{-g}\, g^{\a\be}\,$ as follows (see for instance \cite{landau,mtw}):
\beq
\partial_{\la\mu}H^{\a\la,\be\mu} =
2\chi\underbrace{(-g)(t^{\a\be}_{_L}+T^{\a\be})}_{\dps{{\cal T}^{\a\be}}}
\qquad\Big[\Ra\partial_\a{\cal T}^{\a\be}=0\Big] \ ,
\eeq
where
\beq
H^{\a\la,\be\mu}\equiv (-g)\(g^{\a\be}g^{\la\mu}
-g^{\a\mu}g^{\la\be}\) = \ggot^{\a\be}Ê\ggot^{\la\mu} - \ggot^{\a\mu}Ê\ggot^{\la\be}
\eeq
and $t^{\a\be}_{_L}$ denotes the pseudo-tensor of Einstein--Landau, defined by the following expression:
\beq
\begin{aligned}
&2\chi(-g)t_{_L}^{\a\be} \equiv \partial_\rho \ggot^{\a\be}\partial_\s \ggot^{\rho\s} -\partial_\rho \ggot^{\a\rho}\partial_\s \ggot^{\be\s} +\frac12 g^{\a\be} g_{\la\mu}\partial_\rho\ggot^{\la\s}\partial_\s\ggot^{\mu\rho}
\\[1ex]
&\hspace{2cm} - g_{\la\mu}\partial_\rho\ggot^{\la\s}\big(g^{\a\rho}\partial_\s \ggot^{\be\mu}+ g^{\be\rho}\partial_\s\ggot^{\a\mu}\big)     +g_{\la\mu} g^{\rho\s}\partial_\rho \ggot^{\a\la} \partial_\s \ggot^{\be\mu}
\\[1ex]
&\hspace{2cm}+\frac18\big(g^{\a\be} g^{\rho\s}- 2 g^{\a\rho}g^{\be\s}\big)\big(g_{\la\mu} g_{\tau\nu}- 2 g_{\la\tau}g_{\mu\nu}\big)\partial_\rho \ggot^{\la\mu} \partial_\s \ggot^{\tau\nu} \ ,
\end{aligned}
\eeq

Thorne approach consists firstly on introducing  the deviation of the metric density respect to the Minkowski metric
\beq
\hgot^{\a\be}\equiv \ggot^{\a\be} \!-\eta^{\a\be}  \ ,
\eeq
and then writing the Einstein equation (81) as follows:
\beq
\square\,\hgot^{\a\be} + E_{_G}^{\a\be} = 2\chi{\cal T}^{\a\be} - E_{_H}^{\a\be}\equiv 2\chi{\cal T}_{\rm th}^{\a\be} \ ,
\eeq
 $\square$ being the flat D'Alembert operator, and  the following notation is used:
\beq
\begin{aligned}
&E_{_G}^{\a\be} \equiv \ggot^{\a\be}\partial_{\la\mu}\hgot^{\la\mu}-  \ggot^{\a\la}\partial_{\la\mu}\hgot^{\be\mu} -  \ggot^{\be\la}\partial_{\la\mu}\hgot^{\a\mu}+2 \partial_\la\hgot^{\a\be}\partial_\mu\hgot^{\la\mu} -  \partial_\la\hgot^{\a\la}\partial_\mu\hgot^{\be\mu}
\\[1.5ex]
&E_{_H}^{\a\be} \equiv \hgot^{\la\mu}\partial_{\la\mu}\hgot^{\a\be}-  \partial_\la\hgot^{\a\mu}\partial_\mu\hgot^{\be\la} \ ,
\end{aligned}
\eeq
By demanding  the  coordinates to be harmonic, i.e., they must satisfy $\partial_\la\hgot^{\a\la}= 0$,  it turns out to be the following conclusion obtained:
\beq
E_{_G}^{\a\be} =0  \coma \partial_\a E_{_H}^{\a\be} =0 \ ,
\eeq
and so, Einstein equation (85) are reduced to
\beq
\square\,\hgot^{\a\be}  =  2\chi{\cal T}_{\rm th}^{\a\be} \qquad\Big[\Ra\partial_\a{\cal T}^{\a\be}_{\rm th}=0\Big] \ ,
\eeq
where
\beq
{\cal T}_{\rm th}^{\a\be}\equiv {\cal T}^{\a\be}- \frac1{2\chi}\Big(\hgot^{\la\mu}\partial_{\la\mu}\hgot^{\a\be}- \partial_\la\hgot^{\a\mu}\partial_\mu\hgot^{\be\la}\Big) \ ,
\eeq
which represents an effective energy--momentum pseudo-tensor (containing {\it  ``matter"} and {\it  ``field"}) that is conserved in the ordinary sense \big(let us note that according to (85) and the  condition of harmonic coordinates, both tems in (93) are divergence free\big). Subsequently Thorne defines the {\it static\/}  multipoles as follows:
\beq
M_{\rm th}^{i_1i_2\dots i_n} = \int x^{i_1i_2\dots i_n}\,{\cal T}_{\rm th}^{00} \,d^3\vec{x} = \frac1{2\chi} \int x^{i_1i_2\dots i_n}\,\Delta\hgot^{00} \,d^3\vec{x} 
\eeq
where  integrals are extended to all the space and  $ \Delta $ is the flat Laplacian operator.

\medskip
With this expression of Thorne we can proceed  identically to the classic case of generalized Gauss  theorem and conclude in the following flow integral:
\beq
M_{\rm th}^{i_1i_2\dots i_n}  = \frac1{2\chi} \oint_\infty \Big(x^{i_1i_2\dots i_n}\,\partial_k\hgot^{00} - \hgot^{00}\partial_k x^{i_1i_2\dots i_n}\Big)d\tilde\s^k \ ,
\eeq
having taken into account that this time $\Delta  x^{i_1i_2\dots i_n}=0$. If we restrict ourselves now to the axial symmetry case, then
\beq
\begin{aligned}
M^{\rm th}_n &= \frac{(2n-1)!!}{n!}\,e_{i_1i_2\dots i_n} M_{\rm th}^{i_1i_2\dots i_n}
\\[1.5ex]
&= \frac1{2\chi} \oint_\infty\Big\{r^n P_n(\cos\th)\,\partial_r\hgot^{00} - \hgot^{00}\,\partial_r\big[r^n P_n(\cos\th)\big]\Big\} d\tilde\s \ ,
\end{aligned}
\eeq
which  is virtually identical to (22)  if we substitute the Newtonian potential $ \F $ by the component $ \hgot^{00}$ of the deviation of the metric density. Now considering  (23)   an expression similar  to (24)   is obtained:
\beq
M^{\rm th}_n = \frac18\, r^{n+1}\!\oint_{-1}^{+1} P_n(\cos\th)\,\big(r\partial_r\hgot^{00} -n\,\hgot^{00}\big)\,d(\cos\th) \ ,
\eeq
where once again the circle at the sign of integration means that the limit  $r\to\infty$ must be taken.

\medskip
To calculate the integral (97) we must use the following structure of $ \hgot^{00} $ (obtained from the Appendix):
\beq
\hgot^{00} = -4\sum_{q=0}^\infty\frac{M_q}{r^{q+1}}P_q(\cos\th) + \sum_{q=0}^\infty \frac1{r^{q+2}}X^{(q)}(\cos\th) \ ,
\eeq
where $X^{(q)}(\cos\th)$ represents a polynomial of degree  $q$. From here we obtain the following result:
\beq
r\partial_r\hgot^{00} -n\,\hgot^{00} =  4\sum_{n=0}^\infty
\frac{(q+1+n)M_q}{r^{q+1}}\,P_q(\cos\th) - \sum_{q=0}^\infty \frac{q+2+n}{r^{q+2}}X^{(q)}(\cos\th) \ ,
\eeq
expression that introduced in (97) leads to the desired equality $ M_n^{\rm th} = M_n $, since the second summation results in a zero integral (by reasoning, similar to the classic case,  as in Section 5).

%%%%%%%%%%%%%%%%%%%%%%%%%%%%%%%%%%
%%%%%%%%%%%%%%%%%%%%%%%%%%%%%%%%%%
%%%%%%%%%%%%%%%%%%%%%%%%%%%%%%%%%%

\section{Conclusions}

In this work we have described what we call ``generalized Gauss theorem"(GGT) in classical gravitation, a virtually unknown result in standard bibliography. As we have seen, it can express the multipole moments of the source of the gravitational field as flows throughout any surface containing it. Henceforth  the Gauss theorem,  which allows us to write the mass of the object as the field flow through any envelope surface, is  generalized.

\medskip
We have shown that the definition of  Komar mass (35)  can be understood, at least in the static case, as a relativistic generalization of the Gauss theorem (GT). Indeed, the natural introduction of the mass (46), via the  Einstein  equation (45), can identify it with the mass of Komar by means of  the  flow integral (48).

\medskip
In our judgement the most important aspect of this work is the  generalization of  the previous result. That is, we have defined the static Relativistic Multipole Moments (RMM)  as integrals over all space which  finally identify with the moments of Thorne (or Geroch) by means of flow integrals throughout the surface of infinity. Flow integrals are calculated in associated asymptotically cartesian harmonic polar coordinates, and    the structure of the Weyl metrics in these coordinates, obtained by the authors in previous articles, is used.

\medskip
Finally we have reviewed, in the static case, the approach of the multipole moments of  \,Thorne. As  is known, they are defined as integrals over all the space invol\-ving the  time component of a pseudo-tensor $ T_{\rm th}^{\alpha \beta} $ which is  divergence free, provided that harmonic coordinates are used. As the starting equation possess a Minkowskian  character, it is almost immediate that the application of the generalized Gauss theo\-rem leads to  flow integrals that reproduce the asymptotic multipoles.

%%%%%%%%%%%%%%%%%%%%%%%%%%%%%%%%
%%%%%%%%%%%%%%%%%%%%%%%%%%%%%%%%
%%%%%%%%%%%%%%%%%%%%%%%%%%%%%%%%

\section{Appendix. Approximate Weyl metrics in ``spherical harmonic coordinates" }

Weyl metrics are the most general solution of {\it vacuum  axially symmetric} Einstein equations for the static case. In ``Canonical spherical coordinates" of Weyl $\{t, \tilde r,  \tilde \th, \tilde \vf \}$ these metrics are written as follows:
\beq
ds^2 = -e^{2\Psi}dt^2+ e^{-2\Psi}\Big[e^{2\G}\big(d\tilde r^2 +\tilde r^2\,d\tilde\th^2\big) +\be^2 r^2 \sin^2\th \,d\tilde\vf^2\Big] \ ,
\eeq
 $\be$, $\Psi$ y $\G$, are functions of  $(\tilde r,\tilde\th)$, given by the following expressions:
\beq
\left\{\begin{aligned}
&\be(\tilde r,\tilde\th)=1 \coma \Psi(\tilde r,\tilde\th) = \sum_{n=0}^\infty\frac{a_n}{\tilde r^{\,n+1}}P_n\big(\!\cos\tilde\th\,\big)
\\[1ex]
&\G(\tilde r,\tilde\th) = \sum_{n,k=0}^\infty\frac{(n+1)(k+1)}{n+k+2}\frac{a_n a_k}{\tilde r^{\,n+k+2}} \big(P_{n+1} P_{k+1}-P_n P_k\big)
\end{aligned}\right. \ ,
\eeq
where the argument of the Legendre  polynomials in the second formula is  $ \cos\tilde\th $. We have included the function $ \be = 1 \, $ to indicate that, in general,  inside of the source this function  is not trivial \big(see for instance \cite{kramer_sol,gurle}\big)

\medskip
In \cite {jl_mq} we got the relativistic moments $ M_n $ of these metrics in terms of the Weyl constants $ a_n $ up to  a fairly high order. The procedure used to do that primarily relies on  the definition of Thorne, carrying out the following steps. First of all, spherical coordinates $ \{t, r, \th, \vf \} $ associated to asymptotically cartesian  harmonic ones are determined. Then, the  moments $M_n$ are the coefficients corresponding to  the terms like
\beq
\frac1{r^{\,n+1}}P_n\big(\!\cos\th\,\big)
\eeq
arising in the series expansion of the component $g_{tt}$ of the metric.

\medskip
The calculation of the harmonic coordinates  $\{t, r, \th, \vf\}$  is approximate, so they are provided by  series expansion in the inverse of the radial coordinate $r$ up to a high order. These expressions allow us  to  write the components of the metric up to order  $1/r^9$ (those expressions are shown at the end of this Appendix), and  the line element is taken in the following form:
\beq
ds^2= \g_{tt}dt^2 + \g_{rr}dr^2 + 2\g_{r\th}\,dr\,(rd\th) + \g_{\th\th}(rd\th)^2 + 
 \g_{\vf\vf}(r\sin\th d\vf)^2 \ ,
\eeq
i.e., the  ``euclidean orthonormal co--basis" (60) has been used.

\medskip
As can be seen, the components of the metric have two distinguished kind of terms; the first ones  reproduce the Newtonian multipole expansion and the se\-cond terms (summations involving the notation ${\cal R}^{(q)}$) contain the so-called {\it rests of  Thorne}, with the   following structure:
\beq
\begin{aligned}
&\g_{tt}= -1 + 2\sum_{q=0}^\infty \frac{M_q}{r^{q+1}} P_q(w) + 2\sum_{q=0}^\infty \frac1{r^{q+2}}{\cal R}^{(q)}_{tt}(w)
\\[1ex]
&\g_{ii}= 1 + 2\sum_{q=0}^\infty \frac{M_q}{r^{q+1}} P_q(w) + 2\sum_{q=0}^\infty \frac1{r^{q+2}}{\cal R}^{(q)}_{ii}(w)
\\[1ex]
&\g_{r\th}= \sum_{q=0}^\infty \frac1{r^{q+3}}{\cal R}^{(q)}_{r\th}(w) \ ,
\end{aligned}
\eeq
where $i= \{r, \th, \vf\}$ and $w\!\equiv\!\cos\th$. The rests of Thorne ${\cal R}^{(q)}_{tt}(w)$ and ${\cal R}^{(q)}_{ii}(w)$ are polynomials of degree $q$ in the variable $w$ and they appear in this development  as certain combinations of the Legendre polynomials \big(from Weyl series (101)\big): 
\beq
\begin{aligned}
&{\cal R}^{(q)}_{tt}(w)\equiv \sum_{s=0}^{q} A_{tt}^{qs} P_s(w)
\\[1ex]
&{\cal R}^{(q)}_{ii}(w)\equiv \sum_{s=0}^{q} A_{ii}^{qs} P_s(w)  \ .
\end{aligned}
\eeq
Regarding to the rests ${\cal R}^{(q)}_{r\th}(w)$ they appear as certain combinations of the associated Legendre functions of first kind $P^1_m(w)$
\beq
 {\cal R}^{(q)}_{r\th}(w)\equiv \sum_{s=0}^{q} A_{r\th}^{qs} P^1_{s+1}(w)  \ ,
\eeq
and so they are polynomials of degree $q$ in the variable $w$ multiplied by a factor  $\sin\th$, i.e., 
\beq
 {\cal R}^{(q)}_{r\th}(w)= \sqrt{1-w^2}\, Q^{(q)}_{r\th}(w) \ .
\eeq
Let us note that the absence of odd degree polynomials (as well as associated functions)  is due to the equatorial symmetry considered in our case.

\beq
\begin{aligned}
\hspace{-.3cm}\bullet\ \g_{tt} =-1 &+\frac{2M_0}{r}+\frac{2M_2}{r^3}P_2+\frac{2M_4}{r^5}P_4+\frac{2M_6}{r^7}P_6+\frac{2M_8}{r^9}P_8
\\[1ex]
 &-\frac{2}{r^2}M_0^2 + \frac{2}{r^3}M_0^3-\frac2{r^4}M_0\(M_0^3+2M_2P_2\)
 \\[1ex]
 &+ \frac2{r^5}M_0^2\(M_0^3+\frac{22}7M_2P_2\)
 \\[1ex]
 &- \frac2{r^6}\left[M_0^6+\frac15M_2^2+\frac27M_2\(15M_0^3+M_2\)\!P_2+2\(\frac9{35}M_2^2+M_0 M_4\)\! P_4\right]
  \\[1ex]
 &+ \frac2{r^7}\left[M_0^7+\frac{71}{105}M_0M_2^2 + \frac1{21}M_0M_2\big(115M_0^3 +  17M_2\big)P_2\right. 
 \\[1ex]
 &\hspace{4cm}+\left. \frac3{11}M_0\(\frac{166}{35}M_2^2 + 13M_0 M_4\) P_4\right]
 \\[1ex]
 &- \frac2{r^8}\left[M_0^8+\frac{148}{105}M_0^2M_2^2 + 2M_2\!\(\frac{10}3M_0^5 +  \frac{125}{147}M_0^2M_2+\frac27M_4\!\)\!P_2\right. 
 \\[1ex]
 &\hspace{.5cm}+\left. 2\!\(\frac{3684}{2695}M_0^2M_2^2 +\frac{28}{11}M_0^3 M_4+ \frac{20}{77}M_2 M_4\!\)\! P_4+2\!\(\frac5{11}M_2 M_4+M_0M_6\!\)\! P_6\right]
\\[1ex]
 &+ \frac2{r^9}\left[M_0^9+\frac{5209}{2100}M_0^3M_2^2-\frac{2}{105}M_2^3\right. 
   \\[1ex]
 &\hspace{0.5cm}+M_2\(\frac{260}{33}M_0^6+\frac{468911}{1601700}M_0^3M_2 + \frac{30}{77}M_2^2+ \frac{470}{231}M_0M_4\)P_2 
 \\[1ex]
 &\hspace{0.5cm}+ \(\frac{816479}{175175}M_0^3M_2^2 +\frac{132}{455}M_2^3 + \frac{994}{143}M_0^4 M_4+\frac{1426}{1001}M_0M_2M_4                                                                                                               \) P_4
 \\[1ex]
 &\hspace{0.5cm}+\left.\(\frac27M_2^3+\frac{70}{33}M_0M_2M_4+4M_0^2 M_6\) P_6\right]
\end{aligned}
\eeq
\beq
\begin{aligned}
\hspace{-.5cm}\bullet\ \g_{rr} =\,&1 +\frac{2M_0}{r}+\frac{2M_2}{r^3}P_2+\frac{2M_4}{r^5}P_4+\frac{2M_6}{r^7}P_6+\frac{2M_8}{r^9}P_8
\\[1ex]
 &+\frac{2}{r^2}M_0^2 + \frac{2}{r^3}M_0^3+\frac2{r^4}M_0\(M_0^3+2M_2P_2\)
 \\[1ex]
 &+ \frac2{r^5}M_0^2\(M_0^3+\frac{22}7M_2P_2\)
 \\[1ex]
 &+ \frac2{r^6}\left[M_0^6+\frac1{15}M_2^2+\frac27M_2\!\(\!15M_0^3-\frac13M_2\!\)\!P_2+\(\frac{36}{35}M_2^2+2M_0 M_4\! \)\!P_4\right]
  \\[1ex]
 &+ \frac2{r^7}\left[M_0^7+\frac{43}{105}M_0M_2^2 + \frac1{21}M_0M_2\big(115M_0^3 +  M_2\big)P_2\right. 
 \\[1ex]
 &\hspace{4.3cm}+\left. \frac3{11}M_0\(\frac{298}{35}M_2^2 + 13M_0 M_4\) P_4\right]
 \\[1ex]
 &+ \frac2{r^8}\left[M_0^8+\frac{2623}{2100}M_0^2M_2^2 + \frac{1}{3}M_2\(20M_0^5 +  \frac{6707}{4900}M_0^2M_2+\frac47M_4\)P_2\right. 
 \\[1ex]
 &\hspace{1.2cm}+\(\frac{55771}{13475}M_0^2M_2^2 +\frac{56}{11}M_0^3 M_4- \frac{24}{77}M_2 M_4P_2\) P_4
 \\[1ex]
 &\hspace{5.5cm} +\left. 2\(\frac{35}{33}M_2M_4+M_0 M_6\) P_6\right]
  \\[1ex]
 &+ \frac2{r^9}\left[M_0^9+\frac{1019}{420}M_0^3M_2^2-\frac{6}{35}M_2^3\right. 
   \\[1ex]
 &\hspace{1.2cm}+M_2\(\frac{260}{33}M_0^6+\frac{5419}{4620}M_0^3M_2 + \frac{46}{231}M_2^2+ \frac{14}{11}M_0M_4\)P_2 
 \\[1ex]
 &\hspace{1.2cm}+ \(\frac{32243}{5005}M_0^3M_2^2 +\frac{828}{5005}M_2^3 + \frac{994}{143}M_0^4 M_4+\frac{34}{143}M_0M_2M_4                                                                                                               \) P_4
 \\[1ex]
 &\hspace{1.2cm}+\left.\(\frac{58}{77}M_2^3+\frac{50}{11}M_0M_2M_4+4M_0^2 M_6\) P_6\right]
\end{aligned}
\eeq
\beq
\begin{aligned}
\bullet\ \g_{r\th} = & -\frac2{3r^4}M_0M_2 P_{21} - \frac4{3r^5}M_0^2M_2 P_{21}
\\[1ex]
&+\frac1{r^6}\left[\frac{4}{21}M_2\(-8M_0^3 + M_2\)P_{21} -\frac15\(\frac97M_2^2+2M_0M_4 \)P_{41}\right]
\\[1ex]
&+\frac2{r^7}\left[\frac{2}{7}M_0M_2\(\!-3M_0^3 +\frac13 M_2\!\)\!P_{21} -M_0\(\frac37M_2^2+\frac25M_0M_4\! \)\!P_{41}\right]
\\[1ex]
&+\frac1{r^8}\left[\frac{1}{21}M_2\(\!-38M_0^5 +\frac{943}{700}M_0^2 M_2+\frac43M_4\!\)\!P_{21} \right.
\\[1ex]
&\hspace{2cm}+\(-\frac{15782}{13475}M_0^2M_2^2-\frac{62}{55}M_0^3M_4+\frac{20}{77}M_2M_4\! \)\!P_{41}
\\[1ex]
&\hspace{7cm}\left.-\frac27\(\frac{145}{99}M_2M_4+M_0M_6\!\)\!P_{61}\right]
\\[1ex]
&+\frac1{r^9}\left[-\frac{1}{21}M_2\(\!40M_0^6 +\frac{151}{50}M_0^3 M_2+4M_2^2\!\)\!P_{21} \right.
\\[1ex]
&\hspace{1.3cm}+\frac2{11}\(-\frac{1571}{175}M_0^3M_2^2-\frac{3}{35}M_2^3-8M_0^4M_4+\frac85M_0M_2M_4\! \)\!P_{41}
\\[1ex]
&\hspace{4.5cm}\left.-\frac47\(\frac3{11}M_2^3+\frac{74}{33}M_0M_2M_4+M_0^2M_6\!\)\!P_{61}\right]
\end{aligned}
\eeq
\beq
\begin{aligned}
\hspace{-.8cm}\bullet\ \g_{\th\th} =\,&1 +\frac{2M_0}{r}+\frac{2M_2}{r^3}P_2+\frac{2M_4}{r^5}P_4+\frac{2M_6}{r^7}P_6+\frac{2M_8}{r^9}P_8
\\[1ex]
 &+\frac{1}{r^2}M_0^2 +\frac1{3r^4}M_0M_2(1+5P_2)
+ \frac2{3r^5}M_0^2M_2\(1-\frac{4}7P_2\)
 \\[1ex]
 &+ \frac1{r^6}\left[\frac{13}{21}M_0^3M_2+\frac4{15}M_2^2+\frac2{15}M_0M_4\right.
 \\[1ex]
 &\left.\hspace{1.5cm}-\frac13\!\(\!M_0^3M_2-\frac{19}7M_2^2-2M_0M_4\!\)\!P_2-\frac65\(\frac17M_2^2-M_0 M_4\! \)\!P_4\right]
  \\[1ex]
 &+ \frac2{r^7}\left[\frac27M_0^4M_2+\frac{8}{105}M_0M_2^2 + \frac2{15}M_0^2M_4 \right.
\\[1ex]
 &\qquad\left. -\frac13M_0\(\frac57M_0^3M_2-\frac{17}7M_2^2 -2M_0M_4\!\)\!P_2
-\frac1{11}M_0\(\frac{426}{35}M_2^2 + \frac{14}5M_0 M_4\!\)\! P_4\right]
 \\[1ex]
 &+ \frac1{r^8}\left[\frac{11}{21}M_0^5M_2+\frac{1651}{4620}M_0^2M_2^2 +\frac{52}{165}M_0^3M_4+\frac1{42}M_2M_4+\frac1{14}M_0M_6\right.
 \\[1ex]
 & \hspace{1cm}+\(\!-\frac37M_0^5M_2 +  \frac{76843}{40425}M_0^2M_2^2+\frac{52}{33}M_0^3M_4+\frac{37}{42}M_2M_4+\frac5{14}M_0M_6\!\)\!P_2
 \\[1ex]
 &\hspace{1.2cm}+\(\!-\frac{2952}{1225}M_0^2M_2^2 -\frac45M_0^3 M_4+ \frac{317}{154}M_2 M_4+\frac9{14}M_0M_6\) P_4
 \\[1ex]
 &\hspace{6.8cm} +\left. \(\!-\frac{445}{462}M_2M_4+\frac{13}{14}M_0 M_6\) P_6\right]
\\[1ex]
 &+ \frac1{r^9}\left[\frac{10}{21}M_0^6M_2+\frac{879}{1925}M_0^3M_2^2+\frac2{21}M_2^3 +\frac4{11}M_0^4M_4 +\frac{11}{35}M_0M_2M_4+\frac17M_0^2M_6\right. 
\\[1ex]
 &\hspace{1.5cm}+\(\!-\frac{100}{231}M_0^6M_2+\frac{161383}{80850}M_0^3M_2^2 + \frac4{231}M_2^3+ \frac{20}{11}M_0^4M_4\right.
 \\[1ex]
 & \hspace{8.5cm}\left.+\frac{97}{77}M_0M_2M_4 +\frac57M_0^2M_6\!\)\!P_2 
\\[1ex]
 &\hspace{0.4cm}+ \(\!-\frac{481934}{175175}M_0^3M_2^2 +\frac{300}{1001}M_2^3 - \frac{222}{143}M_0^4 M_4+\frac{17887}{5005}M_0M_2M_4 +\frac97M_0^2M_6\!\)\! P_4
 \\[1ex]
 &\hspace{5.5cm}\left.-\frac17\(\frac{40}{11}M_2^3+\frac{461}{11}M_0M_2M_4+M_0^2 M_6\) P_6\right]
\end{aligned}
\eeq
\beq
\begin{aligned}
\hspace{-.8cm}\bullet\ \g_{\vf\vf} =\,&1 +\frac{2M_0}{r}+\frac{2M_2}{r^3}P_2+\frac{2M_4}{r^5}P_4+\frac{2M_6}{r^7}P_6+\frac{2M_8}{r^9}P_8
\\[1ex]
 &+\frac1{r^2}M_0^2 -\frac1{3r^4}M_0M_2\(1-7P_2\)
 - \frac2{3r^5}M_0^2M_2\(1-\frac{10}7P_2\)
 \\[1ex]
 &+ \frac1{r^6}\left[-\frac{13}{21}M_0^3M_2+\frac25M_2^2-\frac2{15}M_0M_4 \right.
 \\[1ex]
 &\hspace{1.2cm}\left.+\!\(\!\frac{19}{21}M_0^3M_2+\frac37M_2^2-\frac23M_0M_4\!\)\!P_2+\frac25\(\frac37M_2^2+7M_0 M_4\! \)\!P_4\right]
  \\[1ex]
 &+ \frac2{r^7}\left[-\frac27M_0^4M_2+\frac{12}{35}M_0M_2^2 - \frac2{15}M_0^2M_4\right. 
 \\[1ex]
 &\hspace{1.1cm}\left.+M_0\!\(\frac13M_0^3M_2 -\frac17  M_2^2-\frac23M_0M_4\!\)\!P_2 -\frac2{11}M_0\!\(\frac{81}{35}M_2^2 + \frac{37}5M_0 M_4\!\)\! P_4\right]
 \\[1ex]
 &+ \frac2{r^8}\left[-\frac{11}{21}M_0^5M_2+\frac{3227}{3300}M_0^2M_2^2 -\frac{52}{165}M_0^3M_4-\frac1{42}M_2M_4-\frac1{14}M_0M_6\right.
 \\[1ex]
& \hspace{1cm}+\( \frac{13}{21}M_0^5M_2 -  \frac{5039}{16170}M_0^2M_2^2-\frac{52}{33}M_0^3M_4+\frac{43}{42}M_2M_4-\frac5{14}M_0M_6\!\)\!P_2 
 \\[1ex]
 &\hspace{1cm}+\(\!-\frac{2204}{2695}M_0^2M_2^2 +\frac{164}{55}M_0^3 M_4+ \frac9{14}M_2 M_4-\frac9{14}M_0M_6\!\) \!P_4
 \\[1ex]
&\hspace{7.5cm} +\left.\frac1{14}\(5M_2M_4+43M_0 M_6\) P_6\right]
\\[1ex]
 &+ \frac1{r^9}\left[-\frac{10}{21}M_0^6M_2+\frac{6724}{5775}M_0^3M_2^2-\frac2{21}M_2^3 -\frac4{11}M_0^4M_4 -\frac{11}{35}M_0M_2 -\frac17M_0^2M_6\right. 
\\[1ex]
 &\hspace{1cm}+\(\frac{40}{77}M_0^6M_2-\frac{42379}{80850}M_0^3M_2^2 + \frac{16}{77}M_2^3 - \frac{20}{11}M_0^4M_4\right.
 \\[1ex]
&\left. \hspace{7cm}+\frac{17}{11}M_0M_2M_4-\frac57M_0^2M_6\!\)\!P_2 
 \\[1ex]
 &\hspace{1cm}+ \(-\frac{164422}{175175}M_0^3M_2^2 -\frac{12}{1001}M_2^3 + \frac{402}{143}M_0^4 M_4-\frac{51}{65}M_0M_2M_4  -\frac97M_0^2M_6 \!\)\! P_4
 \\[1ex]
 &\hspace{4cm}+\left.\frac17\(\!-\frac{16}{11}M_2^3 - 9M_0M_2M_4 + 29M_0^2 M_6\!\)\! P_6\right]
\end{aligned}
\eeq
%%

%%%%%%%%%%%%%%%%%%%%%%%%%%%%%%%%%%
%%%%%%%%%%%%%%%%%%%%%%%%%%%%%%%%%%
%%%%%%%%%%%%%%%%%%%%%%%%%%%%%%%%%%

\newpage
\section*{Acknowledgments}
This  work  was  partially supported by the Spanish  Ministerio de Econom\'\i a y Competitividad under Research Project No. FIS2015-65140-P (MINECO), and the Consejer\'\i a
de Educaci\'on of the Junta de Castilla y Le\'on under the Research Project Grupo
de Excelencia GR234.

%%%%%%%%%%%%%%%%%%%%%%%%%%%%%%%%%%%%%
%%%%%%%%%%%%%%%%%%%%%%%%%%%%%%%%%%%%%
%%%%%%%%%%%%%%%%%%%%%%%%%%%%%%%%%%%%%


\begin{thebibliography}{88}

\bibitem{geroch} R. Geroch, {\it J. Math. Phys.\/} {\bf Vol. 11}, 1955 (1970).

{\it J. Math. Phys.\/} {\bf Vol. 11}, 2580 (1970)

\bibitem{hansen} R.O. Hansen, {\it J. Math. Phys.\/} {\bf Vol. 15}, 46 (1974).

\bibitem{thorne} K. S. Thorne  {\it Rev. Mod. Phys.}  {\bf Vol. 52, No. 2. Part I}, 299 
(1980).

\bibitem{beigsimon} R. Beig , {\it Gen. Rel. Grav.\/} {\bf Vol. 12}, 439 (1980).

R. Beig and W. Simon , {\it Commun. Math. Phys.\/} {\bf Vol. 78}, 75 (1980).

R. Beig and W. Simon , {\it Acta Phys. Austr.\/} {\bf Vol. 53}, 249 (1981).

R. Beig and W. Simon , {\it Proc. Royal Soc. London\/} {\bf Vol. A 376}, 333 (1981).


\bibitem{quevedo} H. Quevedo {\it Fortschr. Phys.} {\bf 38}, 733 (1990).
H. Quevedo {\it Phys. Rev.} {\bf 33}, 334 (1986).

\bibitem{fhp} Fodor, G., Hoenselaers, C. and Perj\'es Z.,  {\it J. Math. Phys.}, {\bf 30}, 2252, (1989).


\bibitem{bakdal}  B\"ackdahl, T., Herberthson,  M., (2005) {\it Class. Quantum Grav.} {\bf 22}, 3585.

Herberthson,  M., (2004) {\it Class. Quantum Grav.} {\bf 21}, 5121.

B\"ackdahl, T., Herberthson,  M., (2006) {\it Class. Quantum Grav.} {\bf 23}, 5997.

B\"ackdahl, T., Herberthson,  M., (2005) {\it Class. Quantum Grav.} {\bf 22}, 1607.


\bibitem{luisgyrv}
L.Herrera and J.L.Hern\'andez-Pastora, {\it Classical and Quantum Gravity.} {\bf 17} 3617-3625 (2000)

L. Herrera, and Hern\'andez-Pastora
{\it Journal of Math. Phys.} {\bf 41}, 7544. (2000)

\bibitem{luisgeodv}
 Hern\'andez-Pastora, J.L, and Ospino, J. {\it Phys. Rev. D} {\bf 82}, 104001 (14 pages) (2010).
 
 L. Herrera, (2005) {\it Found. Phys. Lett.} {\bf 18}, 21-36.
 
L. Herrera, J. Carot, N. Bolivar and E. Lazo, {\it Int.J.Theor.Phys.} {\bf 48}, 3537-3546, (2009)

J.L.Hern\'andez-Pastora and L. Herrera, {\it Classical and Quantum Gravity}. {\bf 28}, 225026 (2011)

\bibitem{kramer_sol} D. Kramer, H. Stephani, M. MacAllum and E. Herlt, {\it Exact solutions of Einstein field equations\/}, Cambridgge University Press (1980).


\bibitem{gurle} N. G\"urlebeck  {\it Phy. Rev. D}, {\bf 90},  024041 (2014)

\bibitem{komar} A. Komar,  {\it Phys. Rev.}  {\bf Vol. 113, No. 3}, 934 (1959).

\bibitem{tolman} R. C. Tolman,  {\it Phys. Rev. Second Series}  {\bf Vol. 35, No. 8}, 875 (1930).

\bibitem{jl_mq} J.L. Hern\'andez-Pastora, J. Mart\'{\i}n 
{\it Gen. Rel. and Grav..} {\bf 26}, 877 (1994).

\bibitem{jl_tesis} J.L. Hern\'andez-Pastora,
{\it Ph. D. Relativistic gravitational fields close to Schwarzschild solution}. Universidad de Salamanca. (1996).

\bibitem{weinberg} S. Weinberg, {\it Gravitation and Cosmology\/}, Wiley \& Sons, New York (1972).

\bibitem{tulzy} W. Tulczyjew, {\it Acta Phys. Polon.\/} {\bf Vol. 18}, 393 (1959).

\bibitem{belgeroch} L. Bel {\it Gen. Relat. Gravit.}  {\bf Vol. 1}, 337 (1971).
 
R. Geroch, {\it J. Math. Phys.\/} {\bf Vol. 12}, 918 (1971).


\bibitem{belescard} L. Bel and J.C. Escard {\it Acc. Nazionale dei Lince\/}. {\bf Ser VIII, vol XVI, fasc. 6} (1966)

L. Bel  {\it J. Math. Phys.}  {\bf Vol. 10, No. 4}, 337 (1969)

L. Bel  {\it Gen. Relat. Gravt.}  {\bf Vol. 1, No. 8}, 337 (1971)

\bibitem{kerr_newman} R.P. Kerr  and E.T. Newman, {\it J. Math. Phys.\/} {\bf Vol. 6}, 918 (1965).

\bibitem{mmr} J.A. Cabezas, J. Mart\'{\i}n, A. Molina and E. Ruiz {\it Gen. Relat. Gravit.}  {\bf Vol.39}, 707 (2007)

arXiv: gr-qc/0611013 v1, 2 Nov 2006

\bibitem{joseluis} J.L.  Hern\'andez-Pastora,
{\it Class. Quantum Grav.} {\bf 27}, 045006 (2010). 

\bibitem{landau} L. Landau and E. Lifchitz, {\it Th\'eorie des Champs\/}, \'Editions MIR, Moscou (1970).

 L. Landau and E. Lifshitz, {\it The Classical Theory of Fields\/}, Pergamon, Oxford (1971).

\bibitem{mtw} C.W. Misner, K.S. Thorne and J.A. Wheeler (MTW), {\it Gravitation\/}, Freeman, San Francisco (1973).



\end{thebibliography}
\end{document}